\documentclass[aps,prl, reprint,superscriptaddress, preprintnumbers]{revtex4-1}
\usepackage{blindtext}
\usepackage[english]{babel}
\usepackage{amsmath}
\usepackage{subcaption}
\usepackage{feynmp-auto}
\usepackage{mathtools,slashed}
\usepackage{color}
\usepackage{array}
\usepackage{graphicx}
\usepackage[colorlinks=true,citecolor=blue,linktocpage=true,linkcolor=blue]{hyperref}
\usepackage{caption}
\usepackage{float}

 \def\be{\begin{equation}}
\def\ee{\end{equation}}
\def\bea#1\eea{\begin{align}#1\end{align}}

\def\a{\alpha}
\def\b{\beta}

\def\d{\delta}
\def\e{\epsilon}

\def\m{\mu}
\def\n{\nu}

\def\l{\lambda}

\def\r{\rho}

\def\s{\sigma}
\def\e{\epsilon}

\def\bi{\begin{itemize}}
\def\ei{\end{itemize}}

\begin{document}

\preprint{IFT-UAM/CSIC-17-068}
\preprint{FTUAM-17-13}
\preprint{FTI-UCM/201753}
\title{Do the gravitational corrections to the beta functions of the quartic and Yukawa couplings have an intrinsic physical meaning?}

\author{S. Gonzalez-Martin}
\email{sergio.gonzalezm@uam.es}
\affiliation{Instituto de Fisica Teorica UAM/CSIC, C/ Nicolas Cabrera, 13–15, C.U. Cantoblanco, 28049 Madrid, Spain}
\affiliation{Departamento de Fisica Teorica, Universidad Autonoma de Madrid, 20849 Madrid, Spain}

\author{C. P. Martin}
\email{carmelop@fis.ucm.es}
\affiliation{Universidad Complutense de Madrid (UCM), Departamento de Física Teórica I, Facultad de Ciencias Físicas,  Av. Complutense S/N (Ciudad Univ.), 28040 Madrid, Spain}
\begin{abstract}
 We study the beta functions of the quartic  and Yukawa couplings of General Relativity and Unimodular Gravity coupled to the $\lambda\phi^4$ and Yukawa theories with masses. We show that the General Relativity corrections to those beta functions as obtained from the 1PI functional by using the standard MS multiplicative renormalization scheme of Dimensional Regularization are gauge dependent and, further, that they
 can be removed by a non-multiplicative, though local, field redefinition. An analogous analysis is carried out when General Relativity is replaced with Unimodular Gravity. Thus we show that any claim made about the change in the asymptotic behaviour of the quartic and Yukawa couplings made by General Relativity and Unimodular Gravity lack intrinsic physical meaning.
\end{abstract}

\maketitle

\paragraph{Introduction.-}

It is well known that perturbatively quantized general relativity is non-renormalizable due to the mass dimension of the coupling constant $\kappa$ \cite{tHooft}. Moreover, the coupling to matter does not improve this behavior \cite{Deser1,Deser2,Deser3,Deser4}.  However, it can still be treated as an effective field theory well below the scale of the Planck mass $M_P\sim 10^{19}\;\text{GeV}$ \cite{Donoghueeffective1,Donoghueeffective2}.
On the other hand, unimodular gravity is known to be an alternative formulation to General Relativity; it yields the same classical predictions, and moreover it  partially solves the so-called cosmological constant problem \cite{Ellis,Unruh}. However, whether quantum corrections --putting aside the absence of such corrections to the Cosmological Constant in Unimodular Gravity-- will turn Unimodular Gravity into a different theory from General Relativity is an open issue \cite{Alvarezunimodular}. Despite this, it has got the same problems of non-renormalizability that general relativity.

Then, it is clear that both theories should be  regarded as effective field theories. It is in this sense, the asymptotic behaviour of physical effects of quantum general relativity on other fields has been studied. Robinson and Wilczek suggested that when coupled to a Yang-Mills theory, it improves the behaviour of the theory regarding asymptotic freedom \cite{RobinsonWilczek}; but it was proved later  that this result is gauge dependent \cite{Pietrykowski1,Rodigast1}. Further, it is also known that a non-multiplicative renormalization can be used to eliminate some of the contributions to the beta functions in the Yang-Mills case \cite{JohnEllis}.\par

In Reference \cite{Rodigast2}, it was shown that the General Relativity contributions to beta functions of the quartic and Yukawa couplings obtained by using the multiplicative MS scheme of Dimensional Regularization applied to the 1PI functional do not vanish in the de Donder gauge of the graviton field. The contributions obtained lead to asymptotic freedom for appropriate values of the masses involved --among these values are masses of the real Higgs and top quark.  In view of the fate of the corrections found in \cite{RobinsonWilczek}, it is necessary to see whether or not the corrections found in \cite{Rodigast2} are gauge independent.

The first aim of the present paper is to show that the General Relativity corrections to the beta functions of the quartic and Yukawa couplings as computed in the de Donder gauge in \cite{Rodigast2} are gauge dependent artifacts and that, besides, they can be removed by  appropriate non-multiplicative field redefinitions. Thus, we conclude that the General Relativity corrections to the beta functions in question obtained by using the multiplicative MS scheme of Dimensional Regularization applied to the 1PI functional have no intrinsic physical meaning and, that, therefore, any
physical conclusion derived from them cannot be trusted. The second aim is to show that this same situation is reproduced when Unimodular Gravity is used instead of General Relativity. We shall actually see that in the gauge we shall use the Unimodular Gravity corrections to the beta function of the quartic coupling vanish in the Multiplicative MS scheme of Dimensional Regularization.

One word of caution: when, in this paper, we talk about gravity corrections --either from General Relativity or from Unimodular Gravity-- we refer to corrections that are of order $\kappa^2$.

\paragraph{The setting.-} We start from the well known Einstein-Hilbert Lagrangian coupled  to a massive real scalar $\phi$ via a $\phi^4$ interaction and a Dirac fermion $\psi$ via a Yukawa interaction. This is
\bea
 {\cal L_\text{GR}}=&\sqrt{-g}\left\{ -\dfrac{2}{\kappa^2}R+ \bar{\psi}(i\slashed{D}-m_\psi)\psi+\dfrac{1}{2}g^{\m\n}\partial_\m\phi\partial_\n\phi+\nonumber\right.\\
&\left.  -\dfrac{1}{2}m_\phi^2 \phi^2- g\phi\bar{\psi}\psi-\dfrac{\lambda}{4!}\phi^4\right\},
\eea
while for unimodular gravity
\bea{\cal L_\text{UG}}=&-\dfrac{2}{\kappa^2}(-g)^\frac{1}{4}~\left(R+\dfrac{3}{32}\dfrac{\nabla_\m g\nabla^\m g}{g^2}\right)+ \bar{\psi}(i\slashed{D}-m_\psi)\psi+\nonumber\\
&+\dfrac{1}{2}g^{\m\n}\partial_\m\phi\partial_\n\phi-\dfrac{1}{2}m_\phi^2 \phi^2- g\phi\bar{\psi}\psi-\dfrac{\lambda}{4!}\phi^4,\eea
where $\kappa=32\pi G$, and $g,\l$ are -respectively- the Yukawa and the $\phi^4$ coupling constants.

In order to keep explicit the gauge dependence, we use a generalized gauge condition for general relativity:
\be{\cal L}_\text{GR}=\alpha \Big(\partial^\m h_{\m\n}-\dfrac{1}{2}\partial_\n h\Big)^2,\ee
where $\alpha$ is an arbitrary gauge parameter. This yields a propagator
\bea\langle &h_{\m\n}(k)h_{r\s}(-k)\rangle_\text{GR}=\dfrac{i}{2k^2}\left(\eta_{\m\s}\eta_{\n\r}+\eta_{\m\r}\eta_{\n\s}-\eta_{\m\n}\eta_{\r\s}\right)-\\
&-i\left(\dfrac{1}{2}+\a\right)\left(\eta_{\m\r}k_\n k_\s+\eta_{\m\s}k_\m k_\r+\eta_{\n\r}k_\m k_\s+\eta_{\n\s}k_\n k_\r\right)\nonumber.\eea
The gauge fixing and propagator of unimodular gravity are found in \cite{Alvareztrees,Alvarezunimodular} and read
\bea\langle &h_{\m\n}(k)h_{r\s}(-k)\rangle_\text{UG}=\nonumber\\
&=\dfrac{i}{2k^2}\left(\eta_{\m\s}\eta_{\n\r}+\eta_{\m\r}\eta_{\n\s}\right)-\dfrac{i}{k^2}\dfrac{8\a^2-1}{16\a^2 }\eta_{\m\n}\eta_{\r\s}+\nonumber\\&+i\left(\dfrac{k_\r k_\s \eta_{\m\n}}{k^4}+\dfrac{k_\m k_\n \eta_{\r\s}}{k^4}\right)
-4i\dfrac{k_{\m}k_{\n}k_{\r}k_{\s}}{k^6}.
\label{propagatorug}
\eea
Let us remark that in the case of unimodular gravity the interaction comes from $h_{\m\n}\widehat{T}^{\m\n}=\widehat{h}_{\m\n}T^{\m\n}$ with ${T}^{\m\n}$ the energy-momentum tensor and the hat quantities the traceless ones. Therefore one can work with the traceless propagator $\langle \widehat{h}_{\m\n}(k)\widehat{h}_{r\s}(-k)\rangle$ (which can be trivially obtained from \eqref{propagatorug}) and the full energy-momentum tensor, therefore using the same Feynman rules for the vertices as in general relativity, or use \eqref{propagatorug} coupled to $\widehat{T}^{\m\n}$.

\begin{figure}[H]
	\centering
		\begin{subfigure}[t]{0.45\linewidth}
		\begin{fmffile}{scalarprop}
			\begin{fmfgraph*}(100,40)
				\fmfleft{i}
				\fmfright{o}
				\fmf{scalar,tension=3,label=$p$}{v1,i}
				\fmf{scalar,tension=0.7,label=$p-k$}{v2,v1}
				\fmf{scalar,tension=3,label=$p$}{o,v2}
				\fmf{wiggly,left,tension=0.7,label=$k$}{v1,v2}
			\end{fmfgraph*}
		\end{fmffile}
		\caption{$P_S$}
		\label{propscalar}
	\end{subfigure}
	\hfill
	\begin{subfigure}[t]{0.45\linewidth}
		\begin{fmffile}{fermionprop}
			\begin{fmfgraph*}(100,40)
				\fmfleft{i}
				\fmfright{o}
				\fmf{fermion,tension=3,label=$p$}{v1,i}
				\fmf{fermion,tension=0.7,label=$p-k$}{v2,v1}
				\fmf{fermion,tension=3,label=$p$}{o,v2}
				\fmf{wiggly,right,tension=0.7,label=$k$}{v2,v1}
			\end{fmfgraph*}
		\end{fmffile}
		\caption{$P_Y$}
		\label{propfermion}
	\end{subfigure}
	\caption{Corrections to the scalar and fermion propagators.}
\end{figure}
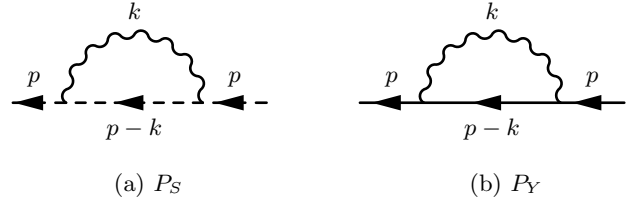

In order to compute the beta functions, the first step is to find the 1PI gravitational corrections to the scalar and fermion propagators. These are shown in figures \ref{propscalar} and \ref{propfermion}.
\vspace{.7cm}

\begin{figure}[!h]
	\centering
	\begin{subfigure}[t]{0.45\linewidth}
	\begin{fmffile}{phi41}
	\begin{fmfgraph*}(120,100)
		\fmfleft{i1,i2}
		\fmfright{o1,o2}
		\fmf{scalar,tension=0.5,label=$p_3$}{i2,v2}
		\fmf{scalar,tension=1,label=$p_1$}{i1,v1}
		\fmf{scalar,tension=1,label=$p_1+k$,l.side=left}{v1,v2}
		\fmf{scalar,tension=0.5,label=$p_4$}{o2,v2}
		\fmf{scalar,tension=1,label=$p_2$,l.side=right}{o1,v3}
		\fmf{scalar,tension=1,label=$p_2-k$,l.side=right}{v3,v2}
		\fmf{wiggly,tension=0,label=$k$,l.side=left}{v3,v1}	
	\end{fmfgraph*}
\end{fmffile}
\caption{+ 5 permutations}
\end{subfigure}
\hfill
\begin{subfigure}[t]{0.45\linewidth}
	\begin{fmffile}{phi42}
	\begin{fmfgraph*}(120,100)
		\fmfleft{i1,i2}
		\fmfright{o1,o2}
		\fmf{scalar,tension=0.5,label=$p_3$}{i2,v2}
		\fmf{scalar,tension=1,label=$p_1$}{i1,v1}
		\fmf{scalar,tension=1,label=$\hspace{-1.5mm}p_1+k$,l.side=right}{v1,v2}
		\fmf{scalar,tension=0.5,label=$p_4$}{o2,v2}
		\fmf{scalar,tension=0.5,label=$p_2$,l.side=right}{o1,v2}
		\fmf{wiggly,left,tension=0,label=$k$,l.side=left}{v1,v2}	
	\end{fmfgraph*}
\end{fmffile}
\caption{+ 4 permutations}
\end{subfigure}
\caption{$\phi^4$ vertices.}
\label{phi4}
\end{figure}
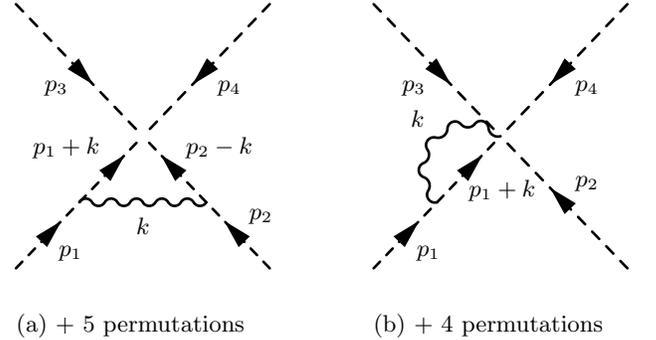

Using the propagators listed above, and computing divergences in dimensional regularization ($D=4+2\e$) these are,
\bea &P_S^{\text{GR}}=\kappa^2\left(-\dfrac{i}{16\pi^2\e}\right)m_\phi^2\Big[1+\Big(\dfrac{1}{2}+\alpha\Big)\Big](p^2-m_\phi^2),\\
&P_S^{\text{UG}}=0,\\ &P_Y^{\text{GR}}=\kappa^2\left(-\dfrac{i}{16\pi^2\e}\right)\left\{\dfrac{3}{8}m_{\psi}p^2-\dfrac{1}{8}p^2\slashed{p}+\dfrac{1}{4}m_\psi^2(\slashed{p}-m_\psi)+\right. \nonumber\\
&\left.+\Big(\dfrac{1}{2}+\alpha\Big)\left[\dfrac{3}{4}m_\psi p^2-\slashed{p}\Big(\dfrac{15}{32}p^2+\dfrac{29}{32}m_\psi^2\Big)-\dfrac{19}{16}m_\psi^3\right]\right\},\\
&P_Y^{\text{UG}}=\kappa^2\left(-\dfrac{i}{16\pi^2\e}\right)\left\{\slashed{p}\Big(\dfrac{3}{16}m_\psi^2-\dfrac{5}{16}p^2\Big)+\dfrac{3}{8}m_\psi p^2\right\}.\eea
The corrections to the $\phi^4$ (1PI) vertex (figure \ref{phi4}) read

\bea V_\phi^\text{GR}&=\kappa^2\l\left(-\dfrac{i}{16\pi^2\e}\right)\Big(\dfrac{3}{2}+\a\Big)\Big[\dfrac{1}{2}\sum_{i=1}^4 p_i^2-4 m_\phi^2\Big],\\
V_\phi^\text{UG}&=0.\eea
Finally, we compute the divergences of the (1PI) Yukawa vertices listed in figure \ref{yukawas}. These ones read
\vspace{.5cm}
\begin{widetext}

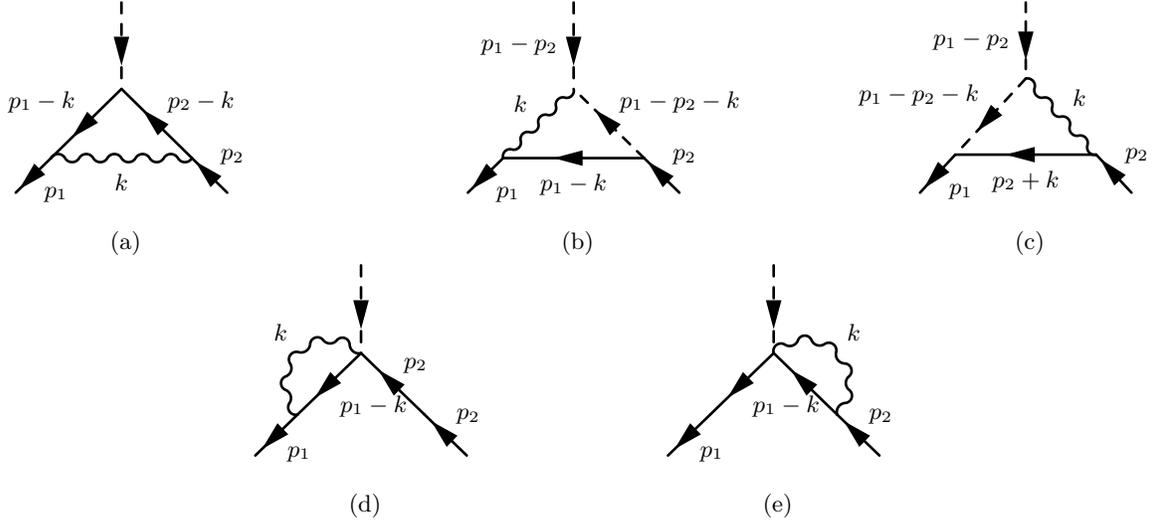
\begin{figure}[]
		\centering
	\vspace{-0.5cm}
		\hfill\begin{subfigure}{0.32\linewidth}
			\centering
		\begin{fmffile}{yukawa1}
			\begin{fmfgraph*}(80,80)
				\fmfbottom{i,o}
				\fmftop{t}
				\fmf{scalar,tension=0.8}{t,v1}
				\fmf{fermion,tension=0.5,label=$p_1-k$,l.side=right}{v1,v3}
				\fmf{fermion,tension=1,label=$p_1$}{v3,i}
				\fmf{fermion,tension=1,label=$p_2$}{o,v2}
				\fmf{fermion,tension=0.5,label=$p_2-k$}{v2,v1}
				\fmf{wiggly,tension=0,label=$k$,l.side=left}{v2,v3}
			\end{fmfgraph*}
		\end{fmffile}
		\caption{}
	\end{subfigure}
	\hfill
	\begin{subfigure}{0.32\linewidth}
		\centering
		\begin{fmffile}{yukawa2}
			\begin{fmfgraph*}(80,80)
				\fmfbottom{i,o}
				\fmftop{t}
				\fmf{scalar,tension=0.8,label=$p_1-p_2$}{t,v1}
				\fmf{wiggly,tension=0.5,label=$k$,l.side=right}{v1,v3}
				\fmf{fermion,tension=1,label=$p_1$}{v3,i}
				\fmf{fermion,tension=1,label=$p_2$}{o,v2}
				\fmf{scalar,tension=0.5,label=$p_1-p_2-k$}{v2,v1}
				\fmf{fermion,tension=0,label=$p_1-k$,l.side=left}{v2,v3}
			\end{fmfgraph*}
		\end{fmffile}
		\caption{}
	\end{subfigure}
	\hfill
	\begin{subfigure}{0.32\linewidth}
		\centering
		\begin{fmffile}{yukawa3}
			\begin{fmfgraph*}(80,80)
				\fmfbottom{i,o}
				\fmftop{t}
				\fmf{scalar,tension=1,label=$p_1-p_2$}{t,v1}
				\fmf{scalar,tension=0.5,label=$p_1-p_2-k$,l.side=right}{v1,v3}
				\fmf{fermion,tension=1,label=$p_1$}{v3,i}
				\fmf{fermion,tension=1,label=$p_2$}{o,v2}
				\fmf{wiggly,tension=0.5,label=$k$,l.side=right}{v2,v1}
				\fmf{fermion,plain,tension=0,label=$p_2+k$,l.side=left}{v2,v3}
			\end{fmfgraph*}
		\end{fmffile}
		\caption{}
	\end{subfigure}
	
	\begin{subfigure}{0.3\linewidth}
		\centering
		\begin{fmffile}{yukawa4}
			\begin{fmfgraph*}(80,80)
				\fmfbottom{i,o}
				\fmftop{t}
				\fmf{scalar,tension=1.4}{t,v1}
				\fmf{fermion,tension=1,label=$p_1-k$,l.side=left}{v1,v3}
				\fmf{fermion,tension=1.5,label=$p_1$}{v3,i}
				\fmf{fermion,tension=1.5,label=$p_2$}{o,v2}
				\fmf{fermion,tension=1,label=$p_2$}{v2,v1}
				\fmf{wiggly,right,tension=0.0,label=$k$,l.side=right}{v1,v3}
			\end{fmfgraph*}
		\end{fmffile}
		\caption{}
	\end{subfigure}
	\begin{subfigure}{0.3\linewidth}
		\centering
		\begin{fmffile}{yukawa5}
			\begin{fmfgraph*}(80,80)
				\fmfbottom{i,o}
				\fmftop{t}
				\fmf{scalar,tension=1.4}{t,v1}
				\fmf{fermion,tension=1,label=$p_1-k$,l.side=left}{v1,v3}
				\fmf{fermion,tension=1.5,label=$p_1$}{v3,i}
				\fmf{fermion,tension=1.5,label=$p_2$}{o,v2}
				\fmf{fermion,tension=1}{v2,v1}
				\fmf{wiggly,right,tension=0.0,label=$k$}{v2,v1}
			\end{fmfgraph*}
		\end{fmffile}
		\caption{}
	\end{subfigure}
	\caption{Contributions to the Yukawa vertex.}
	\label{yukawas}
		\end{figure}

	\bea
	V_\psi^{\text{GR}}&=g\kappa^2 \left(-\dfrac{i}{16\pi^2\e}\right)\Big[ -\dfrac{1}{4}m_\phi^2-\dfrac{3}{4}m_\psi^2+\dfrac{1}{16}(p_1+p_2)^2+\dfrac{1}{4}m_\psi(\slashed{p}_1+\slashed{p}_2)+\dfrac{1}{8}\slashed{p}_1\slashed{p}_2\Big]+\nonumber \\
	&+ g\kappa^2\left(-\dfrac{i}{16\pi^2\e}\right)\Big(\dfrac{1}{2}+\a\Big)\Big[-m_\phi^2-\dfrac{57}{16}m_\psi^2+\dfrac{47}{32}(p_1^2+p_2^2)-\dfrac{13}{8}p_1\cdot p_2 +m_\psi (\slashed{p}_1+\slashed{p}_2)-\dfrac{9}{16}\slashed{p}_1\slashed{p}_2\Big]\\
	V_\psi^{\text{UG}}&=g\kappa^2 \left(-\dfrac{i}{16\pi^2\e}\right)\Big[ \dfrac{9}{16}(p_1^2+p_2^2)-\dfrac{3}{8}p_1\cdot p_2+\dfrac{3}{16}m_\psi(\slashed{p}_1+\slashed{p}_2)-\dfrac{3}{8}\slashed{p}_1\slashed{p}_2\Big] \eea
\end{widetext}
\clearpage

  \paragraph{Beta functions-}  We shall proceed now to the computation of the Yukawa and quartic coupling  beta function gravitational corrections coming from General Relativity and Unimodular Gravity. To use the well known multiplicative MS renormalization scheme of Dimensional Regularization, we define
\bea &g_0=\mu^{-\e}Z_gZ_\psi^{-1}Z_\phi^{-1/2}g, && Z_g=1+\d Z_g,\\
&\Psi_0=Z^{1/2}\Psi, && Z_\Psi,=1+\d Z_\Psi,\\
&\bar{\Psi}_0=Z^{1/2}\bar{\Psi}, && Z_\phi=1+\d Z_g, \\
&m_{\Psi_0}=Z_mZ_\Psi^{-1}m_\Psi, && Z_{m_\Psi}=1+\d Z_{m_\Psi},\\
&m_{\phi_0}=Z_mZ_\phi^{-1}m_\phi, && Z_{m_\phi}=1+\d Z_{m_\phi}.
\eea

\setcounter{table}{3}
\begin{table}[h!]
	\centering
\begin{tabular}{m{3cm} m{1cm} m{3cm}}
	
	\begin{fmffile}{scalarcounter}
	\begin{fmfgraph*}(80,40)
		\fmfleft{i}
		\fmfright{o}
		\fmf{scalar,tension=3,label=$p$}{v1,i}
		\fmf{scalar,tension=3,label=$p$}{o,v1}
		\fmfv{decor.shape=circle,decor.filled=empty,
			decor.size=(.15w)}{v1}
		\fmfv{l=\scalebox{2}{$\mathbf {\times}$},label.dist=0}{v1}
	\end{fmfgraph*}
\end{fmffile}  &\hfill$=$& $i(\d Z_\Psi \slashed{p}-\d Z_{m_\Psi} m_{\Psi})$,\\
\begin{fmffile}{fermioncounter}
	\begin{fmfgraph*}(80,40)
		\fmfleft{i}
		\fmfright{o}
		\fmf{fermion,tension=3,label=$p$}{v1,i}
		\fmf{fermion,tension=3,label=$p$}{o,v1}
		\fmfv{decor.shape=circle,decor.filled=empty,
			decor.size=(.15w)}{v1}
		\fmfv{l=\scalebox{2}{$\mathbf {\times}$},label.dist=0}{v1}
	\end{fmfgraph*}
	\end{fmffile} &\hfill$=$& $i(\d Z_\phi p^2-\d Z_{m_\phi} m_{\phi}^2)$,\\
\begin{fmffile}{yukawacounter}
	\begin{fmfgraph*}(80,90)
		\fmfbottom{i,o}
		\fmftop{t}
		\fmf{scalar,tension=2}{t,v1}
			\fmf{fermion,tension=1,label=$p_1$}{v1,i}
		\fmf{fermion,tension=1,label=$p_2$}{o,v1}
		\fmfv{decor.shape=circle,decor.filled=empty,
			decor.size=(.15w)}{v1}
		\fmfv{l=\scalebox{2}{$\mathbf {\times}$},label.dist=0}{v1}
		
	\end{fmfgraph*}
\end{fmffile}&\hfill$=$& $-ig\m^{-\e}\d Z_g$.
\end{tabular}
\caption{Counterterms.}
\label{counterterms}
\end{table}

The counterterms obtained from the previous definitions are given in figure \ref{counterterms}.

Following the standard MS procedure, the wave function renormalizations ($\d Z_\Psi$ and $\d Z_\Psi$) are obtained by imposing that the contributions proportional to $\slashed{p}$ in the sum given in figure \ref{wfrenorm} are finite as $\e\to 0$. This yields the values

\bea \d Z_\phi&=\dfrac{1}{16\pi^2\e}\kappa^2 m_\phi^2\Big[1+\Big(\dfrac{1}{2}+\a\Big)\Big],\\
\d Z_\Psi&=\dfrac{1}{16\pi^2\e}\kappa^2 m_\Psi^2\Big[\dfrac{1}{4}+\Big(\dfrac{1}{2}+\a\Big)\dfrac{29}{32}\Big].\eea

\begin{table}[h!]
	\centering
	\begin{tabular}{m{3cm} m{1cm} m{3cm}}
		
			\begin{fmffile}{scalarprop2}
			\begin{fmfgraph*}(80,40)
				\fmfleft{i}
				\fmfright{o}
				\fmf{scalar,tension=3,label=$p$}{v1,i}
				\fmf{scalar,tension=0.7}{v2,v1}
				\fmf{scalar,tension=3,label=$p$}{o,v2}
				\fmf{wiggly,right,tension=0.7}{v2,v1}
			\end{fmfgraph*}
		\end{fmffile}& \hfill$+$&	\begin{fmffile}{scalarcounter2}
			\begin{fmfgraph*}(80,40)
				\fmfleft{i}
				\fmfright{o}
				\fmf{scalar,tension=3,label=$p$}{v1,i}
				\fmf{scalar,tension=3,label=$p$}{o,v1}
				\fmfv{decor.shape=circle,decor.filled=empty,
					decor.size=(.15w)}{v1}
				\fmfv{l=\scalebox{2}{$\mathbf {\times}$},label.dist=0}{v1}
			\end{fmfgraph*}	\end{fmffile}\\
	\begin{fmffile}{fermionprop2}
	\begin{fmfgraph*}(80,40)
		\fmfleft{i}
		\fmfright{o}
		\fmf{fermion,tension=3,label=$p$}{v1,i}
		\fmf{fermion,tension=0.7}{v2,v1}
		\fmf{fermion,tension=3,label=$p$}{o,v2}
		\fmf{wiggly,right,tension=0.7}{v2,v1}
	\end{fmfgraph*}
\end{fmffile}& \hfill$+$& 	\begin{fmffile}{fermioncounter2}
\begin{fmfgraph*}(80,40)
	\fmfleft{i}
	\fmfright{o}
	\fmf{fermion,tension=3,label=$p$}{v1,i}
	\fmf{fermion,tension=3,label=$p$}{o,v1}
	\fmfv{decor.shape=circle,decor.filled=empty,
		decor.size=(.15w)}{v1}
	\fmfv{l=\scalebox{2}{$\mathbf {\times}$},label.dist=0}{v1}

\end{fmfgraph*}
\end{fmffile}
\end{tabular}
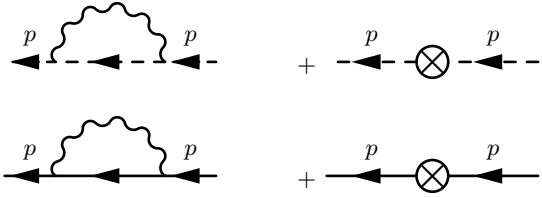
\caption{Wave function renormalization.}
\label{wfrenorm}
\end{table}

For $\d Z_g$, we demand that there is no singularity independent of the external momenta at $\e\to 0$ in the sum of figure \ref{dZg}; hence

\be \d Z_g=\dfrac{1}{16\pi^2\e}\kappa^2 \Big\{m_\phi^2\Big[\dfrac{1}{4}+\Big(\dfrac{1}{2}+\a\Big)\Big]+m_\Psi^2\Big[\dfrac{3}{4}\Big(\dfrac{1}{2}+\a\Big)\dfrac{57}{16}\Big]\Big\}\ee

\begin{table}[ht!]
	\centering
	\begin{tabular}{m{3cm} m{1cm} m{3cm}}
		
	\begin{fmffile}{yukawaloop}
		\begin{fmfgraph*}(80,90)
			\fmfbottom{i,o}
			\fmftop{t}
			\fmf{scalar,tension=2}{t,v1}
			\fmf{fermion,tension=1}{v1,i}
				\fmf{fermion,tension=1}{o,v1}
				\fmfblob{0.8cm}{v1}
					
		\end{fmfgraph*}
	\end{fmffile}&\centering $+$&\begin{fmffile}{yukawacounter5}
		\begin{fmfgraph*}(80,90)
			\fmfbottom{i,o}
			\fmftop{t}
			\fmf{scalar,tension=2}{t,v1}
			\fmf{fermion,tension=1}{v1,i}
			\fmf{fermion,tension=1}{o,v1}
			\fmfv{decor.shape=circle,decor.filled=empty,
				decor.size=(.15w)}{v1}
			\fmfv{l=\scalebox{2}{$\mathbf {\times}$},label.dist=0}{v1}
			\end{fmfgraph*}
	\end{fmffile}
	\end{tabular}
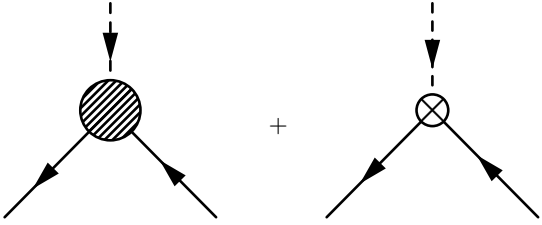
	\caption{Yukawa vertex renormalization.}
		\label{dZg}\vspace{-0.4cm}
\end{table}

Defining $ \b_g=\m \dfrac{dg(\m)}{d\m}$,
and using  standard techniques, one obtains the General Relativity contribution, $\b_g^\text{GR}$, to $\b_g$, at order $\kappa^2$, from $\d \widetilde{Z}_g=\d Z_g-\d Z_\Psi-\dfrac{1}{2}\d Z_\phi$:
\pagebreak
\be \b_g^\text{GR}=\dfrac{1}{16\pi^2}\kappa^2\Big\{m_\phi^2\Big[\dfrac{1}{2}-\Big(\dfrac{1}{2}+\a\Big)\Big]+m_\Psi^2\Big[-1-\Big(\dfrac{1}{2}+\a\Big)\dfrac{85}{16}\Big].\label{bggrm}\ee

The explicit dependence on the parameter $\a$ shows the gauge-dependent nature of this beta function in presence of gravity. Insofar as no physical observables can depend on the gauge, no physical consequences should be extracted from here.

We follow the same procedure for unimodular gravity to find
\bea
&\d Z_\Psi^\text{UG}=\dfrac{1}{16\pi^2\e} \kappa^2m_\Psi^2\dfrac{3}{16},\\
	&\d Z_\phi^\text{UG}=0,\\
	&\d Z_g^\text{UG}=0,
\eea
so that
\be \b_g^\text{UG}=\dfrac{1}{16\pi^2}\kappa^2 m_\Psi^2\dfrac{3}{16}.\label{bgugm}\ee
We can see that we get a difference between general relativity and unimodular gravity by comparing \eqref{bggrm} and \eqref{bgugm}. However we will see in the sequel that we can get rid of these beta functions by using a \textit{non-multiplicative} renormalization, this is, by performing a field redefinition.
\begin{widetext}
Let us now define
 \bea &g_0=\mu^{-\e}Z_gZ_\psi^{-1}Z_\phi^{-1/2}g, &\phi_0&=\phi+\dfrac{1}{2}\d Z_\phi \phi,\\
  &\Psi_0=\Psi+\dfrac{1}{2}\d Z_\Psi\Psi+\dfrac{1}{2}a_1\kappa^2m_\Psi^2\phi\Psi+\dfrac{1}{2}b_1 \kappa^2 m_\phi^2 \phi \Psi,&m_{\Psi_0}&=(1+\d Z_{m_\Psi})m_\Psi ,\\
&\bar{\Psi}_0=\bar{\Psi} +\dfrac{1}{2}a_1\kappa^2 m_\Psi^2\bar{\Psi}\phi+\dfrac{1}{2}b_1 \kappa^2 m_\phi^2 \bar{\Psi}\phi, &m_{\phi_0}&=(1+\d Z_{m_\phi})m_\phi .
\eea
Therefore the matter lagrangian can be written as
\bea & \bar{\Psi}_0(i\slashed{\partial}-m_{\Psi_0})\Psi_0+\dfrac{1}{2}(\partial_\m \phi_0 \partial^\m \phi_0-m_{\phi_0}^2\phi_0^2)-g_0 \bar{\Psi_0 \phi_0 \Psi_0}= \bar{\Psi}(i\slashed{\partial}-m_{\Psi})\Psi+\dfrac{1}{2}(\partial_\m \phi \partial^\m \phi-m_{\phi}^2\phi^2)-g\mu^{-\e} \bar{\Psi \phi \Psi}+\nonumber\\
&+\{\d Z_{\Psi}\bar{\Psi}i\slashed{\partial}\Psi+\d Z_\phi\partial_\m\phi\partial^\m\phi-m_\Psi\d Z_{m_\Psi}\bar{\Psi}\Psi-\dfrac{1}{2}m_\phi^2\d Z_{m_\phi}\}-g\m^{-\e}\{\d Z_g+a_1\kappa^2m_\Psi^2+b_1\kappa^2 m_\phi^2\}\bar{\Psi}\phi\Psi+\nonumber\\
&+\dfrac{1}{2}(a_1\kappa^2m_{\Psi}^2+b_1\kappa^2 m_\phi^2)[i\bar{\Psi}\phi\slashed{\partial}\Psi+i\slashed{\partial}(\phi\Psi)].\eea
\end{widetext}

While the counterterms for the scalar and fermion field propagator remain unchanged with respect to the multiplicative renormalization, the counterterm for the vertex is now given by the expression in figure \ref{nonmultcounterterm}.

\begin{table}[]
	\centering
	\begin{tabular}{m{3cm} m{.3cm} m{5cm}}
		
		\begin{fmffile}{yukawacounter2}
			\begin{fmfgraph*}(80,90)
				\fmfbottom{i,o}
				\fmftop{t}
				\fmf{scalar,tension=2}{t,v1}
				\fmf{fermion,tension=1}{v1,i}
				\fmf{fermion,tension=1}{o,v1}
				\fmfv{decor.shape=circle,decor.filled=empty,
					decor.size=(.15w)}{v1}
				\fmfv{l=\scalebox{2}{$\mathbf {\times}$},label.dist=0}{v1}
			\end{fmfgraph*}
		\end{fmffile}&$=$& $-ig(\m^{-\e}\d Z_g+a_1\kappa^2 m_\Psi^2+b_1\kappa^2 m_\phi^2)$.
	\end{tabular}
	\caption{Counterterm.}
\label{nonmultcounterterm}
\end{table}
Imposing again that the sum in figure \ref{dZg} is zero (plus terms depending on the external momenta) when $\e\to 0$, we find\pagebreak
\bea \d Z_\phi&=\dfrac{1}{16\pi^2\e}\kappa^2 m_\phi^2\Big[1+\Big(\dfrac{1}{2}+\a\Big)\Big],\\
\d Z_\Psi&=\dfrac{1}{16\pi^2\e}\kappa^2 m_\Psi^2\Big[\dfrac{1}{4}+\dfrac{29}{32}\Big(\dfrac{1}{2}+\a\Big)\Big],\\
\d\widetilde{Z}_g&=\d Z_g-\d Z_\Psi-\dfrac{1}{2}\d Z_{\phi}=\nonumber\\
&=\dfrac{1}{16\pi^2\e}\kappa^2m_\phi^2\Big[-\dfrac{1}{4}+\dfrac{1}{2}\Big(\dfrac{1}{2}+\a\Big)\Big]+\nonumber\\
&+\dfrac{1}{16\pi^2\e}\kappa^2m_\Psi^2\Big[\dfrac{1}{2}+\dfrac{85}{32}\Big(\dfrac{1}{2}+\a\Big)\Big]-\nonumber\\
&-a_1\kappa^2m_\Psi^2-b_1\kappa^2 m_\phi^2.\eea

It is clear that by choosing
\bea
a_1&=\dfrac{1}{16\pi^2\e}\Big[\dfrac{1}{2}+\dfrac{85}{32}\Big(\dfrac{1}{2}+\a\Big)\Big],\\
b_1&=\dfrac{1}{16\pi^2\e}\Big[-\dfrac{1}{4}+\dfrac{1}{2}\Big(\dfrac{1}{2}+\a\Big)\Big],\eea
we shall wipe out the gravitational correction to $\d\widetilde{Z}_g$ so the gravitational corrections to the beta function of the Yukawa coupling is now given by

\be \b_g^\text{GR}\Big|_\text{gravitational}=0.\ee

We have seen here that the gravitational contribution to $\b_g$ can be brushed away by carrying out a field redefinition. Therefore, it is  an \textit{inessential} \cite{Weinberg:1980gg} contribution. However, notice that one cannot do the same with the contributions in absence of gravity, which show that they are \textit{essential} contributions.

Finally, we can perform the same non-multiplicative renormalization for unimodular gravity finding
\bea \d Z_\phi^\text{UG}&=0,\\
\d Z_\Psi^\text{UG}&=\dfrac{1}{16\pi^2\e}\kappa^2 m_\Psi^2\dfrac{3}{16},\\
\d\widetilde{Z}_g^\text{UG}&=\d Z_g^\text{UG}-\d Z_\Psi^\text{UG}-\dfrac{1}{2}\d Z_{phi}^\text{UG}=\nonumber\\
=&-\dfrac{1}{16\pi^2\e}\kappa^2m_\Psi^2\dfrac{3}{16}-\nonumber\\
&-a_1\kappa^2m_\Psi^2-b_1\kappa^2m_\phi^2.\eea

Accordingly, we can set
\bea
&a_1=-\dfrac{1}{16\pi^2\e}\dfrac{3}{16},\\
&b_1=0,\eea

to make again $ \d\widetilde{Z}_g=0$ and

\be\b_g^\text{UG}\Big|_\text{gravitational}=0.\ee

The computation of the gravitational corrections of the beta function of the $\phi^4$ interaction is done by following an akin the process.  Defining

\be \lambda_0=\lambda \mu^{-2\e}Z_\l Z_\phi^{-2},\quad Z_\l=1+\delta Z_\l, \ee
we have obtained

\bea \d Z_\l^\text{GR}&=\dfrac{4}{16\pi^2\e}\Big(\dfrac{3}{2}+\a\Big),\\
\d Z_\l^\text{UG}&=0.\eea

Hence, one can compute the gravitational corrections to the beta functions of the quartic coupling, $\lambda\phi^4$, to be

\bea \b_\l^\text{GR}&=-\dfrac{1}{4\pi^2}\kappa^2m_\phi^2\Big(\dfrac{3}{2}+\a\Big)\l,\\
\b_\l^\text{UG}&=0.\eea

In this case the beta function of unimodular gravity is directly zero for this particular gauge. For general relativity, as we did with the Yukawa coupling, we can reabsorb this discrepancy by means of a non-multiplicative renormalization. In this case, we can carry out the following field redefinition
\be \phi_0=\phi+\omega_1\phi+\omega_2\kappa^2\partial^2\phi+w_3 \kappa^2\m^{-2\e}\phi^3\d Z_\phi \phi,\ee
and we can set $\b_\l^\text{GR}=0$ by choosing

\bea \omega_1&=-\dfrac{1}{16\pi^2\e}\kappa^2 m_\phi^2,\\
\omega_2&=0,\\
\omega_3&=\dfrac{1}{16\pi^2\e}\dfrac{1}{4!}2\l.\eea

\paragraph{Summary and final discussion.- } In this paper, we have  computed the General Relativity corrections to the beta functions of the Yukawa and $\lambda\phi^4$ theory as obtained from the 1PI functional
by using the standard multiplicative MS dimensional regularization scheme. We have shown that they are gauge dependent and that, besides, they can be set to zero by appropriate, non-multiplicative, field redefinitions: they are {\it inessential} corrections \cite{Weinberg:1980gg}. We thus conclude that these corrections do not have any intrinsic physical meaning and, that, therefore, the statements about asymptotic freedom made in reference \cite{Rodigast2} are not physically meaningful. Of course, the gauge dependence of the gravitational corrections to the beta function can be avoided by using the DeWitt-Vilkovisky action instead of the 1PI functional --as done  in reference \cite{Pietrykowski:2012nc} for the $\lambda\phi^4$ theory--, but, it is plain that those gauge-independent contributions can still be removed by  appropriate non-mutiplicative, but local,  field redefinitions such as the ones --with different coefficients, of course-- introduced in this paper. The use of the DeWitt-Vilkovisky effective action does not give the gravitational corrections in  question any intrinsic physical meaning, so that any conclusion drawn from them also lack intrinsic physical content.

For the sake of comparison, we have carried out a similar computation for the case of Unimodular Gravity --for a gauge-fixing choice which yields no free parameters: the computations are hard enough already-- and found that the corresponding gravitational corrections to the beta functions do not agree with those from General Relativity --curiously enough the corrections to the beta function of the $\lambda\phi^4$ vanish for Unimodular Gravity, and, that they can also be set to zero by appropriate local non-multiplicative field redefinitions. So one cannot use these gravitational corrections to the beta functions in question to distinguish between General Relativity and Unimodular Gravity. In fact, they behave in the same manner from the physical point of view: they are not {\it essential} in either case, for they correspond to field redefinitions.

A couple of final comments are in order. First, we would like to point out that our conclusions are quite in keeping with the conclusions reached in reference \cite{Anber:2010uj} in the massless case, but our approach to the problem is not the same and, besides, our theories are massive. Notice that the contributions computed in the present paper --and in \cite{Rodigast2, Pietrykowski:2012nc}--  vanish if the masses are sent to zero. Secondly, the results that we have presented  are to be taken into account unavoidable when developing the asymptotic safety program as applied to Gravity interacting with matter, with the proviso that the UV divergences computed in this paper correspond to logarithmic divergences when a cutoff is used.


We are grateful to E. Alvarez for illuminating discussions.
This work has received funding from the European Unions Horizon 2020 research and innovation programme under the Marie Sklodowska-Curie grants agreement No 674896 and No 690575. We also have been partially supported by by the Spanish MINECO through grants
FPA2014-54154-P and FPA2016-78645-P, COST actions MP1405 (Quantum Structure of Spacetime) and  COST MP1210 (The string theory Universe).  S.G-M acknowledge the support of the Spanish MINECO {\em Centro de Excelencia Severo Ochoa} Programme under grant  SEV-2012-0249.

\end{document}